\documentclass[letterpaper, 10 pt, conference]{ieeeconf}  

\IEEEoverridecommandlockouts

\title{\LARGE \bf
Polytopic Input Constraints in Learning-Based Optimal Control\\ Using Neural Networks
}

\author{Lukas Markolf and Olaf Stursberg 
\thanks{The authors are with the Control and System Theory Group, Dept. of Electrical Engineering and Computer Science, University of Kassel, Germany.
{\tt\footnotesize \{lukas.markolf, stursberg\}@uni-kassel.de}} 
}

\usepackage{import}
\usepackage{pgfplots}
\pgfplotsset{compat=newest}
\pgfplotsset{plot coordinates/math parser=false}
\usetikzlibrary{arrows.meta}
\usepackage{tikz}
\usetikzlibrary{patterns}    
\usetikzlibrary{arrows}
\usetikzlibrary{positioning}
\usetikzlibrary{calc}
\usepackage{subcaption}
\usepackage{booktabs}
\usepackage{amsmath}
\usepackage{amssymb}
\usepackage{bm}
\usepackage{xcolor}
\usepackage{multirow}
\usepackage{amsfonts}
\usepackage{comment}
\usepackage{fancyhdr}

\newcommand{\softmaxComp}[2]{\text{softmax}_{#1}\left(#2\right)}
\newcommand{\uHat}[1]{\hat{u}^{\left(#1\right)}}

\newtheorem{Problem}{Problem}

\begin{document}

\maketitle

\pubid{\begin{minipage}{\textwidth}\ \\[65pt]
\fbox{\begin{minipage}{\textwidth}
  \textbf{This version has been accepted for publication in Proc. of the 2021 European Control Conference (ECC).} \\
  \copyright2021~European~Control~Association~(EUCA). Personal use of this material is permitted. Permission from EUCA must be obtained for all other uses, in any current or future media, including reprinting/republishing this material for advertising or promotional purposes, creating new collective works, for resale or redistribution to servers or lists, or reuse of any copyrighted component of this work in other works.
  \end{minipage}}
\end{minipage}} 

\newlength\fheight
\newlength\fwidth


\begin{abstract}
This work considers artificial feed-forward neural networks as parametric approximators in optimal control of discrete-time systems. Two different approaches are introduced to take polytopic input constraints into account. The first approach determines (sub-)optimal inputs by the application of gradient methods. Closed-form expressions for the gradient of general neural networks with respect to their inputs are derived. The approach allows to consider state-dependent input constraints, as well as to ensure the satisfaction of state constraints by exploiting recursive reachable set computations. The second approach makes use of neural networks with softmax output units to map states into parameters, which determine (sub-)optimal inputs by a convex combination of the vertices of the input constraint set. The application of both approaches in model predictive control is discussed, and results obtained for a numerical example are used for illustration.
\end{abstract}


\section{INTRODUCTION}

This paper considers discrete-time systems of the type:
\begin{align}\label{eq:dyn}
x_{k+1} = f(x_k, u_k), \quad k \in\{0, 1, \ldots, N-1\},
\end{align}
with time index $k$ and a finite time horizon $N$. The state and input vectors are constrained to $x_k \in X \subset \mathbb{R}^{n_x}$, $k\in\{0,\ldots,N\}$ and $u_k \in U \subset \mathbb{R}^{n_u}$,  $k\in\{0,\ldots,N-1\}$. A standard problem in open-loop optimal control is considered, which is specified next.

\subsection{An Open-Loop Optimal Control Problem}
Let $X_N \subseteq X$ be a target set and $\lbrace X_0, X_1, \ldots, X_N \rbrace$ a sequence of sets with:
\begin{align}
X_k = \lbrace x_k \in X \, \vert \, \exists u_k \in U \text{ with } f(x_k, u_k) \in X_{k+1} \rbrace.
\end{align}
These sets may result from recursive reachable set computations, see e.g. \cite{B99}. Provided that $X_0$ is nonempty, there exists for each initial state $x_0 \in X_0$ at least one input sequence $\lbrace u_0, \ldots, u_{N-1} \rbrace$ that steers $x_0$ into the target set $X_N$, while satisfying the intermediate state and input constraints. Such an admissible input sequence satisfies:
\begin{align} \label{eq:AdmissibleInputSet}
u_k \in U_k(x_k) = \lbrace u_k \in U \, \vert \, f(x_k, u_k) \in X_{k+1} \rbrace.
\end{align}
Let:
\begin{align} 
J(x_k; u_k, \ldots, u_{N-1}) = g_N(x_N) + \sum_{m=k}^{N-1} g_m(x_m, u_m)
\end{align}
be the total costs associated to an input sequence $\lbrace u_k, \ldots, u_{N-1} \rbrace$ for a given state $x_k$. A standard problem in open-loop optimal control considered in this work is then to find the optimal admissible input sequence $\lbrace u_0^{\ast}, \ldots, u_{N-1}^{\ast} \rbrace$ for a given initial state $x_0 \in X_0$ to obtain the optimal costs:
\begin{align} \label{eq:OptimalControlProblem}
J^{\ast}(x_0) = \min_{ \substack{ u_k \in U_k(x_k) \\ k \in\{ 0, \ldots, N-1\} } } J(x_0; u_0, \ldots, u_{N-1}).
\end{align}

\subsection{Approximate Dynamic Programming}
Dynamic programming (DP)~\cite{B57} is known to have a wide range of applicability in optimal control~\cite{B17} and may be used to address the optimal control problem. Starting from:
\begin{align}
J_N^{\ast}(x_N) := g_N(x_N),
\end{align} 
the DP algorithm proceeds backward in time from $N-1$ to $0$ to compute the optimal cost-to-go functions:
\begin{align}
J_k^{\ast}(x_k) = \min_{ u_k \in U_k(x_k) } \Bigl[ g_k(x_k, u_k) + J_{k+1}^{\ast}(f(x_k, u_k)) \Bigr].
\end{align}
It then holds that:
\begin{align}
J_k^{\ast}(x_k) = \min_{ \substack{ u_m \in U_m(x_m) \\ m \in\{ k, \ldots, N-1 \}} } J(x_k; u_k, \ldots, u_{N-1}),
\end{align}
and thus $J^{\ast}(x_0) = J_{0}^{\ast}(x_0)$. Provided that the optimal cost-to-go values are known for all relevant $x_k$ and $k$, the optimal input sequence for $x_0 \in X_0$ is constructed in a forward manner by:
\begin{align} \label{eq:OptSeqForwardManner}
u_k^{\ast} \in \arg\min_{ u_k \in U_k( x_k^{\ast} ) } \Bigl[ g_k( x_k^{\ast}, u_k ) + J_{k+1}^{\ast}\left( f( x_k^{\ast}, u_k ) \right) \Bigl],
\end{align}
with $x_0^{\ast} = x_0$ and $x_{k+1}^{\ast} = f(x_k^{\ast}, u_k^{\ast})$.

Only for simple cases, the DP algorithm leads to closed-form expressions for $J_k^{\ast}$ and for the respective optimal policy $\pi^{\ast}= \lbrace \mu_{1}^{\ast}(\cdot), \ldots, \mu_{N-1}^{\ast}(\cdot) \rbrace$ with $\mu_k^{\ast} : X \rightarrow U$ -- thus, numeric solution is necessary. Unfortunately, numeric DP solutions are known to suffer from the ``curse of dimensionality'', limiting their practical application. Approaches of approximation offer to diminish this problem, and are instrumental to methods of reinforcement learning~\cite{SB18}, approximate/adaptive dynamic programming~\cite{LV09}, or neuro-dynamic programming~\cite{BT96}. 

In~\cite{B19}, two general types of approximative DP-based approaches have been distinguished, namely approximation in value space and approximation in policy space. The approximation of an optimal cost-to-go function $J_k^{\ast}$ by a parametric function $\tilde{J}_k$ is an example for approximation in value space. An approximated optimal input sequence is then obtained in a forward manner similar to~\eqref{eq:OptSeqForwardManner}, with the difference that $J_{k+1}^{\ast}$ is replaced by $\tilde{J}_{k+1}$ for each $k \in \lbrace 0, \ldots, N-2 \rbrace$. On the other hand, the approximation of an optimal policy $\pi^{\ast}$ by an approximating one $\tilde{\pi} = \lbrace \tilde{\mu}_{1}(\cdot), \ldots, \tilde{\mu}_{N-1}(\cdot) \rbrace$, consisting of parametric functions $\tilde{\mu}_k$, is an example for approximation in policy space. 

The use of neural networks as parametric approximator is appealing due to the universal approximation theorem~\cite{C89},~\cite{HSW89}, and the recent success of deep learning~\cite{GBC16}. However, neural networks are hard to analyze, caused by their nonlinear and large-scale structure~\cite{FMP20}. Hence, neural networks are often considered as black-box models, which may be inappropriate in control tasks that require safety guarantees. 

\subsection{Problem Formulation and Contribution }
This work focuses on neural networks as parametric approximators for optimal cost-to-go functions or policies, where the main question investigated here is how to obtain approximately optimal control inputs while taking polytopic input constraints into account. For this, two different problems are considered, and for each of them a solution approach is proposed. 

\textbf{Formulation of Problem~1.} For the case that the optimal cost-to-go functions are approximated by neural networks, Problem~1 is addressed to obtain an input $\tilde{u}_k \in U$ for a given state $x_k$.
\begin{Problem}
Solve:
\begin{align} \label{eq:Problem1}
\min_{ u_k \in \tilde{U}_k( x_k ) } \Bigl[ g_k( x_k, u_k ) + \tilde{J}_{k+1}\left( f( x_k, u_k ), r_{k+1} \right) \Bigl]
\end{align}
for the case that:
\begin{itemize}
\item[1)] $\tilde{U}_k(x_k) \subseteq U$ is a polytope that depends on $x_k$.
\item[2)] $g_k : \mathbb{R}^{n_x} \times \mathbb{R}^{n_u} \rightarrow \mathbb{R}$ and $f$ are functions that are continuously differentiable with respect to $u_k$.
\item[3)] $\tilde{J}_{k+1} : \mathbb{R}^{n_x} \times \mathbb{R}^{n_r} \rightarrow \mathbb{R}$ is a feed-forward neural network with parameter vector $r_{k+1}$, linear output unit, and continuously differentiable activation functions.
\end{itemize}
\end{Problem}
Concerning a continuous input set, nonlinear programming techniques have been suggested in~\cite{B19} in order to address~\eqref{eq:Problem1} for a general parametric approximator $\tilde{J}_{k+1}$, arguing that these techniques may be more efficient than choosing a discretization-based approach. However, details about the application of nonlinear programming for the case that $\tilde{J}_{k+1}$ is established as neural network are not provided. The contribution of the present work is to derive and use a closed-form expression for the gradient of a neural network $\tilde{J}_{k+1}$ with respect to its input vector. Using these expressions, well-known gradient methods can be employed to solve \eqref{eq:Problem1} with polytopic input constraints. As will be seen, an advantage of this approach is that state-dependent input constraints can be considered, and this also enables to ensure the satisfaction of state constraints, even if the gradient method is stopped before a local optimum is found.

\textbf{Formulation of Problem~2.} If the aforementioned approach is not fast enough for real-time application, it may be more promising to establish an approximate optimal policy $\tilde{\pi}$ with neural networks $\tilde{\mu}_k$ mapping states $x_k$ directly into control inputs $\tilde{u}_k$. However, it is typically a challenging task to determine an output set of a neural network for a considered set of network inputs -- see e.g.~\cite{FMP20} for a method to compute an outer approximation of the true output set. In contrast to an a-posteriori analysis, Problem~2 considers the problem of finding a neural network architecture that guarantees the satisfaction of polytopic input constraints a-priori.
\begin{Problem}
Specify a neural network architecture $\tilde{\mu}_k$ as parametric function of $(x_k, r_k)$ which can a-priori guarantee that $\tilde{\mu}_k(x_k, r_k) \in U$ for each $(x_k, r_k) \in \mathbb{R}^{n_x} \times \mathbb{R}^{n_r}$, provided that $U$ is a polytope.
\end{Problem}
It has been shown that the satisfaction of box constraints can be ensured a-priori by properties of common activation functions. In~\cite{FMP20}, e.g., rectified linear units are used for this purpose. This paper proposes an alternative and new approach, allowing to design a controller which ensures general polytopic input constraints. The idea is based on the fact that each element in a polytope can be described as a convex combination of its vertices. In contrast to map the state directly into a control input, this work proposes to use a neural network to map the state into parameters which are used, subsequently, in convex combination to lead to the control input.

The paper is structured such that Sec.~\ref{sec:2} covers the type of considered neural networks and their training. In Sec.~\ref{sec:3}, the solution approaches for Problem~1 and Problem~2 are proposed. Section~\ref{sec:4} presents the application of the proposed approaches in model predictive control and numeric results, before the paper is concluded in Sec.~\ref{sec:5}.
\section{NEURAL NETWORKS} \label{sec:2}
\subsection{Network Structure}
This work focuses on networks with overall mapping defined by a chain structure of the form~\cite{GBC16}:
\begin{align} \label{eq:NNOverallMapping}
h(x) = (h^{(L)} \circ \cdots \circ h^{(2)} \circ h^{(1)})(x),
\end{align}
with layers $h^{(\ell)}$, $\ell\in\{1,\ldots,L\}$.  The final layer $h^{(L)}$ is usually denoted as output layer, while the others are referred to as hidden layers. Let $\eta^{(\ell)}$ denote the output of layer $\ell$, and $\eta^{(0)}$ the input of the overall network:
\begin{align}
\eta^{(0)}(x) &= x, \\
\eta^{(\ell)}(x) &= (h^{(\ell)} \circ \cdots \circ h^{(1)})(x), \quad \ell \in\{ 1, \ldots, L\}.
\end{align} 
The hidden layers are functions of the form:
\begin{align}
h^{(\ell)}(\eta^{(\ell-1)}) = (\phi^{(\ell)} \circ \psi^{(\ell)})(\eta^{(\ell-1)}), \quad \ell \in\{ 1, \ldots, L-1\},
\end{align}
with $\psi^{(\ell)}$ and $\phi^{(\ell)}$ constituting affine and nonlinear transformations, respectively. The affine transformation $\psi^{(\ell)}$ is defined by:
\begin{align}
\psi^{(\ell)}(\eta^{(\ell-1)}) = W^{(\ell)} \eta^{(\ell-1)} + b^{(\ell)}, \quad \ell \in \lbrace 1, \ldots, L-1 \rbrace,
\end{align}  
and is affected by the choice of the weight matrix $W^{(\ell)}$ and the bias vector $b^{(\ell)}$.

Each layer consists of parallel units, each of which defining a vector-to-scalar function. Let $S^{(\ell)}$ be an integer describing the number of units in layer $\ell$. The function of unit $i$ in layer $\ell$ is the $i$-th component of $h^{(\ell)}$. In the case of hidden layers, $h_i^{(\ell)}(\eta^{(\ell-1)}) = \phi_{i}^{(\ell)}(W^{(\ell)} \eta^{(\ell-1)} + b^{(\ell)})$, where $\phi_{i}^{(\ell)}$ is the activation function, often chosen as rectified linear unit or sigmoid function. In this work, linear and softmax output units are considered. For linear output units, the function $h^{(L)}$ is specified as affine transformation:
\begin{align} \label{eq:LinearOutputUnit}
\psi^{(L)}(\eta^{(L-1)}) = W^{(L)} \eta^{(L-1)} + b^{(L)}.
\end{align} 
An affine transformation of type~\eqref{eq:LinearOutputUnit} arises also in softmax output units, where $h_i^{(L)}$ is set to:
\begin{align} \label{eq:SoftmaxOutputUnit}
\softmaxComp{i}{\psi^{(L)}\left(\eta^{(L-1)}\right)} = \frac{\exp\left(\psi_i^{(L)}\left(\eta^{(L-1)}\right)\right)}{\sum_{j=1}^{S^{(L)}}\exp\left(\psi_j^{(\ell)}\left(\eta^{(L-1)}\right)\right)}.
\end{align}

\subsection{Training of the Network}
Consider a parametric class of policies $\tilde{\mu}_k(x_k, r_k)$, where $r_k$ is a parameter vector. In~\cite{B19}, a general scheme for parametric approximation in policy space is given as follows:
\begin{itemize}
\item[1)] introduce a parametric family of policies $\tilde{\mu}_k(x_k, r_k)$;
\item[2)] obtain a large number of state-input samples $(x_k^s, u_k^s)$, $s \in\{ 1, \ldots, q_k\}$, such that $u_k^s$ is a ``good'' input in state $x_k^s$ for any $s$;
\item[3)] determine $r_k$ by solving the regression problem:
\begin{align} \label{eq:RegressionProblem}
\min_{r_k} \sum_{s=1}^{q_k} \Vert u_k^s - \tilde{\mu}_k(x_k^s, r_k) \Vert^{2}.
\end{align}
\end{itemize}
A neural network can be used as a parametric family of policies $\tilde{\mu}_k$, where the elements of the weight matrices and bias vectors are contained in:
\begin{align}
r_k = \begin{bmatrix}
W_{1,1}^{(1)} & \ldots & W_{S^{(L)},S^{(L-1)}}^{(L)} & b_1^{(1)} & \ldots & b_{S^{(L)}}^{(L)}
\end{bmatrix}^{T}.
\end{align}
Analogously, state-cost pairs $(x_k^s, \beta_k^s)$, $s \in \lbrace 1, \ldots, q_k \rbrace$ are chosen to train neural networks approximating cost-to-go functions. Note that the cost function of problem~\eqref{eq:RegressionProblem} is nonconvex, but state-of-the-art methods of nonlinear programming often find sufficiently good solutions. 

To get the training pairs, these may be specified by a human expert, they may originate from solving optimization problems for selected initial states~\cite{MES20}, or from approximate dynamic programming. Sequential dynamic programming is a possible scheme to train approximators for cost-to-go functions~\cite{B19}. Sec.~\ref{sec:4} will show that the solution proposed for Problem~1 is not only relevant for determining control inputs, but also for generating training data by sequential dynamic programming.  
\section{SOLUTION APPROACHES} \label{sec:3}
\subsection{Solution Approach for Problem 1}
Consider a neural network $\tilde{J}_{k+1}$ approximating the optimal cost-to-go function $J_{k+1}^{\ast}$. Once $\tilde{J}_{k+1}$ is trained with state-cost pairs $(x_{k+1}^s, \beta_{k+1}^s)$, $s \in \lbrace 1, \ldots, q_{k+1} \rbrace$, the parameter vector $r_{k+1}$ is obtained and fixed. In order to compute $\tilde{u}_k$ for a state $x_k$ and a parameter vector $r_{k+1}$, the constrained optimization problem~1 is solved:
\begin{align}
\begin{split}
&\min_{u_k} Q_k(u_k) \\
\text{s.t.\ } & u_k \in \tilde{U}_k,
\end{split}
\end{align}
with $Q_k(u_k) := g_k(x_k, u_k) + \tilde{J}_{k+1}( f(x_k, u_k), r_{k+1} )$ and $\tilde{U}_k := \tilde{U}_k(x_k)$. This problem can be solved by gradient methods, for which the handling of the convex constraints $u_k \in \tilde{U}_k$ is standard. The particular step to be highlighted here is the computation of the gradient:
\begin{align}
\begin{split}
\nabla Q_k(u) = & \nabla_{u} g_k(x_k, u) + \\
& \nabla_{u} f(x_k, u) \nabla_f \tilde{J}_{k+1}(f(x_k, u), r_{k+1}).
\end{split}
\end{align}
This step is challenging due to the nonlinear and large-scale structure of the neural network $\tilde{J}_{k+1}$. Fortunately, a closed-form expression for $\nabla_f \tilde{J}_{k+1}$ can be derived, as will be proposed next.

The overall mapping of the neural network $\tilde{J}_{k+1}$ is defined by a chain structure of the form~\eqref{eq:NNOverallMapping}, i.e. $\tilde{J}_{k+1}(\cdot, r_{k+1}) = h(\cdot) = (h^{(L)} \circ \cdots \circ h^{(2)} \circ h^{(1)})(\cdot)$. In Problem~1, continuously differentiable activation functions $\phi_i^{(\ell)}$ are considered, allowing to compute $[\partial h^{(\ell)}/\partial \eta^{(\ell-1)}](\eta^{(\ell-1)}(x))$ for each $\ell \in \lbrace 1, \ldots, L-1 \rbrace$ by using the chain rule:
\begin{align}
\begin{split}
& \frac{\partial h^{(\ell)}(\eta^{(\ell-1)}(x))}{\partial \eta^{(\ell-1)}} \\
& = \frac{\partial \phi^{(\ell)}(\psi^{(\ell)}(\eta^{(\ell-1)}(x)))}{\partial \psi^{(\ell)}}\cdot \frac{\partial \psi^{(\ell)}(\eta^{(\ell-1)}(x)))}{\partial \eta^{(\ell-1)}} \\
&= \frac{\partial \phi^{(\ell)}(\psi^{(\ell)}(\eta^{(\ell-1)}(x)))}{\partial \psi^{(\ell)}}\cdot  W^{(\ell)}, \quad \ell \in\{ 1, \ldots, L-1\}.
\end{split}
\end{align}
Due to the linear output unit (see Problem 1), one gets:
\begin{align}
\frac{\partial h^{(L)}(\eta^{(L-1)}(x))}{\partial \eta^{(L-1)}} = W^{(L)}.
\end{align}
Again with the chain rule, the partial derivative of the overall mapping $h$ with respect to its input vector is computed to:
\begin{align}
\frac{\partial h(x)}{\partial x} = \prod_{i=0}^{L-1} \frac{\partial h^{(L-i)}(\eta^{(L-(i+1))}(x))}{\partial \eta^{(L-(i+1))}},
\end{align}
such that $\nabla_f \tilde{J}_{k+1}$ is obtained in closed-form:
\begin{align}
\nabla_f \tilde{J}_{k+1}(f(x_k, u_k), r_{k+1}) = \left( \left. \frac{\partial h(x)}{\partial x} \right\vert_{x=f(x_k, u_k)} \right)^{T}.
\end{align}

An example for a continuously differentiable activation function is the hyperbolic tangent:
\begin{align}
\tanh(\xi) &= \frac{\exp(\xi)-\exp(-\xi)}{\exp(\xi)+\exp(-\xi)}, \\
\frac{\partial \tanh(\xi)}{\partial \xi} &= 1 - \tanh^2(\xi).
\end{align}
If $\phi_i^{(\ell)} = \tanh$ is chosen as activation function for each unit in the hidden layers, the partial derivative of the overall mapping of the neural network with respect to its input vector can be derived in closed-form to:
\begin{align}
\begin{split}
& \frac{\partial h(x)}{\partial x} = W^{(L)} \prod_{i=1}^{L-1} \Biggl[ \text{diag} \biggr[ \\
& 1 - \tanh^2\left(\psi_j^{(L-i)}\left(\eta^{\left(L-(i+1)\right)}(x)\right)\right) \biggl]_{j=1}^{S^{(L-i)}} W^{(L-i)} \Biggr],
\end{split}
\end{align}
where $\text{diag}$ denotes a diagonal matrix.

Note that partial derivatives of neural networks with respect to the input vector have also been used in other work, see e.g.~\cite{NT99}, but in different context than controller synthesis.

\subsection{Solution Approach for Problem 2}
Problem 2 uses the assumption of a polytopic set $U$. Alternatively to the $\mathcal{H}$-representation (i.e. as set of inequalities), the polytope $U$ can be written in  $\mathcal{V}$-representation as convex hull of its $n_v$ vertices $\lbrace \uHat{1}, \ldots, \uHat{n_v} \rbrace \in U$:
\begin{align}
U = \left\lbrace \sum_{i=1}^{n_v} \lambda_{i} \uHat{i} \, \vert \, \lambda_i \geq 0, \sum_{i=1}^{n_v} \lambda_i = 1 \right\rbrace.
\end{align}
Hence, any $u_k \in U$ can be expressed as convex combination:
\begin{align*}
u_k = \sum_{i=1}^{n_v} \lambda_{i}(u_k) \uHat{i}, \quad \lambda_i(u_k) \geq 0, \quad \sum_{i=1}^{n_v} \lambda_i(u_k) = 1.
\end{align*}
The work in \cite{WS+07} addresses principles of computing the coordinates  $\lambda_i(u_k)$. Here, the proposal is to define the parametric functions $\tilde{\mu}_k$ as a convex combination of the vertices $\{\uHat{1}, \ldots, \uHat{n_v}\}$ of $U$:
\begin{align} \label{eq:ControlLawSol2}
\tilde{\mu}_k(x_k, r_k) = \sum_{i=1}^{n_v} \Lambda_{k, i}(x_k, r_k) \uHat{i},
\end{align}
where $\Lambda_k$ is a neural network with:
\begin{align} \label{eq:LambdaCondition}
&\Lambda_{k, i}(x_k, r_k) \geq 0,\ 
\sum_{i=1}^{n_v} \Lambda_{k, i}(x_k, r_k) = 1 
\end{align}
for all $x_k$ and $r_k$. To ensure that \eqref{eq:LambdaCondition} holds, the use of
softmax output units~\eqref{eq:SoftmaxOutputUnit} is helpful (as done similarly in classification tasks).

As mentioned above, for a neural network which maps states directly onto control inputs, the parameter vector can be obtained by solving \eqref{eq:RegressionProblem}, if pairs $(x_k^s, u_k^s)$, $s \in \lbrace 1, \ldots, q_k \rbrace$ are used as training data. For the procedure in this subsection, however, the state-input pairs are replaced by state-parameter pairs $(x_k^s, \lambda_k^s)$ with the parameter vector $\lambda_k^s$
chosen to yield:
\begin{align}
u_k^s = \sum_{i=1}^{n_v} \lambda_{k, i}^s \uHat{i}.
\end{align}
The parameters of the neural network $\Lambda_k$ are then determined by solving the problem:
\begin{align}
\min_{r_k} \sum_{s=1}^{q_k} \Vert \lambda_k^s - \Lambda_k(x_k^s, r_k) \Vert^{2},
\end{align}
leading to the parameter vector $r_k$.
\section{APPLICATION TO MODEL PREDICTIVE CONTROL} \label{sec:4}
While the solution approaches proposed before are applicable to nonlinear systems \eqref{eq:dyn}, consider for the purpose of illustration 
the linear system:
\begin{align}
f(x_k, u_k) = A x_k + B u_k
\end{align}
subject to polytopic state constraints:
\begin{align}
x_k \in X = \lbrace x_k \in \mathbb{R}^{n_x} \, \vert \, H^{X} x_k \leq h^{X} \rbrace
\end{align}
and polytopic input constraints:
\begin{align}
u_k \in U = \lbrace u_k \in \mathbb{R}^{n_u} \, \vert \, H^{U} u_k \leq h^{U} \rbrace.
\end{align}
Let the target set $X_N \subseteq X$ be a nonempty and control invariant polytope. Then, the overall sequence $\lbrace X_0, \ldots, X_N \rbrace$ consists also of polytopes:
\begin{align}
X_k = \lbrace x \in \mathbb{R}^{n_x} \, \vert \, H^{X_k} x \leq h^{X_k} \rbrace, \quad k \in \lbrace 0, \ldots, N \rbrace,
\end{align}
with $X_i \supseteq X_j$ for $i < j$, $i \in \{0, \ldots N-1\}$~\cite[Theorem 11.2]{BBM17}. Details about the recursive computation of $\lbrace X_0, \ldots, X_N \rbrace$ can also be found in~\cite{BBM17}. 

Model predictive control (MPC) solves a problem of type \eqref{eq:OptimalControlProblem} on-line to determine the $u_k$, but it may not be applicable if the computational effort for online optimization is too large for a timing prescribed by fast system dynamics. Recent work has thus suggested to use neural networks to approximate the MPC control law, e.g. in~\cite{CSA+18, HKT+18, PM20, KL20}. The work in this section shows how the two approaches proposed in Sec.~\ref{sec:3} can be employed to approximate the cost-to-go function of MPC, or the MPC control law respectively, offline by neural networks.  It is assumed for the remainder of this section that the stage cost functions $g_k$, $k \in \lbrace 0, \ldots, N-1 \rbrace$ in~\eqref{eq:OptimalControlProblem} are continuously differentiable with respect to $u_k$.

The solution approach for Problem~1 can be used to approximate the MPC control law by solving:
\begin{align} \label{eq:MPCProb1}
\min_{ u_k \in U_0( x_k ) } \Bigl[ g_0( x_k, u_k ) + \tilde{J}_{1}\left( f( x_k, u_k ), r_{1} \right) \Bigl]
\end{align}
for the current state $x_k \in X_0$. Here, $U_0(x_k)$ is a state-dependent set as defined in~\eqref{eq:AdmissibleInputSet}, and $\tilde{J}_{1}$ refers to a neural network approximating the optimal cost-to-go function $J_1^{\ast}$. For the considered problem setup, the sets $U_k(x_k)$ defined in~\eqref{eq:AdmissibleInputSet} are polytopes, and given by:
\begin{align}
U_k(x_k) = \lbrace u_k \in \mathbb{R}^{n_u} \, \vert \, H^{U_k} u_k \leq h^{U_k}(x_k) \rbrace
\end{align}
for $k \in \lbrace 0 \ldots, N-1 \rbrace$ and with:
\begin{align}
H^{U_k} =\begin{bmatrix}
H^{X_{k+1}} B \\
H^{U}
\end{bmatrix}, \quad
h^{U_k}(x_k) = \begin{bmatrix}
h^{X_{k+1}} - H^{X_{k+1}} A x_k \\
h^{U}
\end{bmatrix}.
\end{align}
Note that the control inputs $\tilde{u}_k$ obtained by this approach ensure the satisfaction of the state and input constraints, even if the gradient method stops before the optimum is reached (to limit computation times).

The neural network $\tilde{J}_{1}$ can be determined by approximation based on sequential dynamic programming~\cite{B19} in a recursive manner, starting with $\tilde{J}_N(x_N, r_N) := g_N(x_N)$. For any $k$, state-cost pairs $(x_k^s, \beta_k^s)$, $s \in \lbrace 1, \ldots, q_k \rbrace$ are generated to determine the parameter vector $r_k$ of $\tilde{J}_k$. Along this line, the space $X_k$ is sampled to generate the states $x_k^s$. The solution approach proposed to solve Problem~1 is then used for each $x_k^s$ to search for an input $\tilde{u}_k^s \in U_k(x_k^s)$ which minimizes $Q_k(u_k) = g_k(x_k^s, u_k) + \tilde{J}_{k+1}(f(x_k^s, u_k), r_{k+1})$. Again, the gradient method may be stopped before the optimum is reached, if necessary to limit the computation time. Once $\tilde{u}_k^s$ is obtained, the state-cost pair $(x_k^s, \beta_k^s)$ follows from $\beta_k^s = g_k(x_k^s, \tilde{u}_k^s) + \tilde{J}_{k+1}(f(x_k^s, \tilde{u}_k^s), r_{k+1})$. The possibility of taking state-dependent input constraints into account is an advantage to be emphasized. These constraints are used here to ensure that $f(x_k^s, \tilde{u}_k^s) \in X_{k+1}$. Poor approximation results of $\tilde{J}_{k+1}$ are most likely for states that do not belong to the space $X_{k+1}$ sampled for training $\tilde{J}_{k+1}$. This leads to approximation errors which may propagate in sequential dynamic programming, and are hence to be avoided. 

Alternatively, a control law of the form~\eqref{eq:ControlLawSol2} can be considered to approximate the MPC control law:
\begin{align} \label{eq:MPCProb2}
\tilde{\mu}_{\text{MPC}}(x_k, r) = \sum_{i = 1}^{n_v} \Lambda_{i}(x_k, r) \uHat{i},
\end{align}
where $\Lambda$ is a neural network with softmax output units, as proposed in the approach for Problem~2. Inputs $u^s$, $s \in \lbrace 1, \ldots, q \rbrace$ may be generated off-line by solving \eqref{eq:MPCProb1} for a large number of states $x^s \in X_0$ in order to obtain a training set consisting of state-input pairs $(x^s, u^s)$. This set can be transformed into a training set consisting of state-parameter pairs $(x^s, \lambda^s)$, see Sec.~\ref{sec:3}, which is then used to determine the parameter vector $r$ of the neural network $\Lambda$ by training. 

For a numeric example, consider the parameterization:
\begin{align*}
A &= \begin{bmatrix}
1.5 & 0 \\ 1 & -1.5
\end{bmatrix}, \quad
B = \begin{bmatrix}
1 & 0 \\ 0 & 1
\end{bmatrix}, \\
H^X &= \begin{bmatrix}
1 & 0 & -1 & 0 \\
0 & 1 & 0 & -1
\end{bmatrix}^T,\
h^X = \begin{bmatrix}
10 & 10 & 10 & 10
\end{bmatrix}^T, \\
H^U &= \begin{bmatrix}
1 & 0 & -1 & 0 & 1 & -1 & -1 & 1 \\
0 & 1 & 0 & -1 & 1 & 1 & -1 & -1
\end{bmatrix}^T, \\
h^U &= \begin{bmatrix}
5 & 5 & 5 & 5 & 7 & 7 & 7 & 7
\end{bmatrix}^T,
\end{align*}
and a target set $X_N=\{0\}$ being just the origin.  The numerical experiment is intentionally chosen simple for illustration, by selecting a short horizon of $N = 6$, a terminal cost $g_N(x_N)=0$, and quadratic stage costs for $k \in\{ 0, \ldots, N-1 \}$:
\begin{align*}
g_k(x_k, u_k) = x_k^T \begin{bmatrix}
1 & 0 \\ 0 & 1
\end{bmatrix} x_k + u_k^T \begin{bmatrix}
1 & 0 \\ 0 & 1
\end{bmatrix} u_k.
\end{align*}
For this quadratic cost function, the constraints imposed by $U_k(x_k)$ are linear, and since the dynamics is linear, too, the optimization problem is convex and can be solved efficiently. This enables to evaluate the performance of the proposed approaches by comparison. (Note again that the proposed approaches are not limited to convex cost functions but can be extended to more general cases provided that the costs per stage are continuously differentiable with respect to the control input.)

\begin{figure*}[t]
\setlength\fheight{7.3cm} 
\setlength\fwidth{10.95cm}
\centering
\begin{minipage}{\textwidth}
\centering
\input{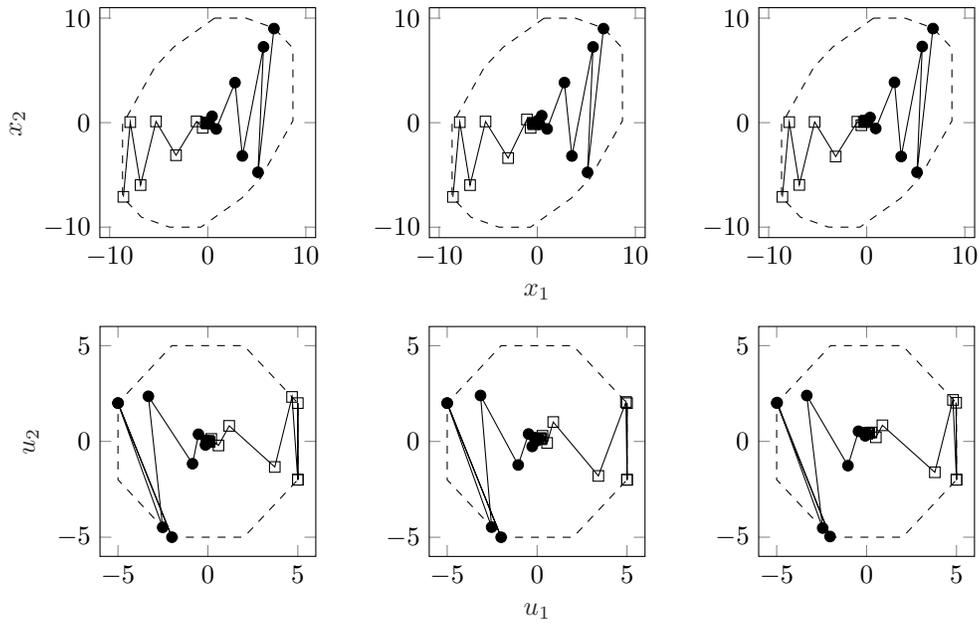}
\caption{State and input trajectories for initial states $[ 6.75 \quad 9 ]^{T}$ and $[ -8.6 \quad -7.1 ]^{T}$. Approximations of the optimal solutions obtained from the proposed methods: top row -- subset of $X$ (dashed line marks $X_0$), bottom row -- $U$ marked by the dashed line. Left column: optimal states and inputs obtained from standard MPC; central column: states and inputs from the approach for Problem~1; right column: trajectories determined by the approach for Problem~2.}
\label{fig:Sequences}
\end{minipage}
\end{figure*}

For both, $\tilde{J}_k$, $k \in \lbrace 1, \ldots, N-1 \rbrace$ and $\Lambda$, a neural network  with one hidden layer consisting of $50$ hidden units with hyperbolic tangent activation functions has been chosen for approximation. For each network, $1000$ training pairs have been generated on the basis of sequential dynamic programming. The approach proposed for Problem 1 was applied by using a conditional gradient method and providing the gradients derived in Sec.~\ref{sec:3}, stopping after $10$ iterations each. On a common notebook (Intel\textsuperscript{\textregistered} Core\texttrademark ~ i$5-7200$U Processor), the computations for these iterations took $0.03$ sec in average, and the linear program arising in the conditional gradient method has been solved by the CPLEXLP solver from the IBM\textsuperscript{\textregistered} ILOG\textsuperscript{\textregistered} CPLEX\textsuperscript{\textregistered} Optimization Studio. The neural networks have been trained using the MATLAB\textsuperscript{\textregistered} Deep Learning Toolbox\texttrademark~with the Levenberg-Marquardt training algorithm~\cite{HM94}. For the determination of optimal MPC inputs, the solver CPLEXQP has been used.

In Fig.~\ref{fig:Sequences}, two vertices of the set $X_0$ are considered as initial states: first, a standard MPC scheme has been used to steer the initial states close to the origin, where the MPC inputs have been obtained from optimal solutions of the open-loop optimal control problem~\eqref{eq:OptimalControlProblem}. The corresponding state and input sequences are shown in the left column of Fig.~\ref{fig:Sequences}. In the second scenario, the inputs have been determined by solving \eqref{eq:MPCProb1} by the approach for Problem~1, and the resulting state and input sequences are presented in the central of the figure. Since the applied gradient method has been stopped after $10$ iterations in the network training, the computation time of an input took around $0.03$ seconds. Eventually, the inputs have been obtained from the control law~\eqref{eq:MPCProb2} in the third scenario. The mapping of the states to the inputs was much faster than the generation of the inputs in scenario 2 (by a factor of $100$). The corresponding state and input sequences can be found in the right column of Fig.~\ref{fig:Sequences}. Even though approximation errors are propagated in the sequential dynamic programming scheme, the results for the two approaches are very similar to the optimal ones. In addition to the satisfaction of the input constraints, the approach considered in scenario 2 has the advantage of ensuring that the state stays within $X_0$. The approach in scenario 3 is much faster, but the determination of an invariant set has not been considered so far.
\section{Conclusion} \label{sec:5}
This paper has proposed two different approaches for considering polytopic input constraints in DP-based control with neural networks. In the first approach, the search for an input minimizing a function represented by a neural network plays an important role, where the input is constrained to a polytopic set. For this, a closed-form expression for the gradient has been derived, allowing to take convex constraints into account in a straightforward way. The advantage is that it is possible to consider state-dependent input constraints, allowing to ensure that the state is steered into a target set while satisfying intermediate state and input constraints. A disadvantage shared with nonlinear programming methods is that typically only local optimal solutions can be determined. Another aspect discussed in this work is that it may be necessary to stop the gradient method before a local optimal solution is found, due to limited computation time -- this, however, does not constitute a fundamental problem, but may result in sub-optimal solutions only. 

The second approach proposed the determination of inputs by convex combinations of state-dependent parameters using the vertices of the polytopic input constraints. It has been shown that neural networks with softmax output units (as state-dependent parameters) satisfy the requirements necessary to ensure the satisfaction of the input constraints. An advantage of this approach is that the generation of the control inputs only requires function evaluations and is therefore promising for on-line application.

Future work will consider state constraints as well as other types of input constraints rather than polytopic ones.


\bibliographystyle{IEEEtran}
\bibliography{ms}

\end{document}